\documentstyle[aas2pp4]{article}

\newcommand{\sn}[1]{$\times 10^{{#1}}$}

\begin{document}

\title{Dust in Starburst Galaxies}

\author{Karl D.\ Gordon}
\affil{Ritter Astrophysical Research Center, The University of Toledo,
   Toledo, OH 43606}
\author{Daniela Calzetti}
\affil{Space Telescope Science Institute, 
   3700 San Martin Dr., Baltimore, MD 21218}
\and
\author{Adolf N.\ Witt}
\affil{Ritter Astrophysical Research Center, The University of Toledo,
   Toledo, OH 43606}

\lefthead{Gordon, Calzetti, \& Witt}
\righthead{Dust in Starbursts}

\begin{abstract}
  To investigate the nature of starbursts' dust, we constructed a
model of the stars and dust in starburst galaxies and applied it to 30
observed starburst spectral energy distributions (SEDs).  The
starburst model was constructed by combining two stellar evolutionary
synthesis models with a model describing the radiative transfer of
stellar photons through dust.  The stellar evolutionary synthesis
models were used to compute the dust-free SEDs for stellar populations
with ages between $1 \times 10^6$ and $15 \times 10^{9}$~years.  Using
a Monte Carlo radiative transfer model, the effects of dust were
computed for average Milky Way (MW) and Small Magellanic Cloud (SMC)
dust, two different star/dust geometries, and locally homogeneous or
clumpy dust.  Using color-color plots, the starburst model was used to
interpret the behavior of 30 starbursts with aperture-matched UV and
optical SEDs (and IR for 19 of the 30) from previous studies.  From
the color-color plots, it was evident that the dust in starbursts has
an extinction curve lacking a 2175~\AA\ bump, like the SMC curve, and
a steep far-UV rise, intermediate between the MW and SMC curves.  The
star/dust geometry which is able to explain the distribution of the 30
starbursts in various color-color plots has an inner dust-free sphere
of stars surrounded by an outer star-free shell of clumpy dust.  When
combined with other work from the literature on the Orion region and
the 30~Dor region of the Large Magellanic Cloud, this work implies a
trend in dust properties with star formation intensity.
\end{abstract}

\keywords{dust, extinction --- galaxies: ISM --- galaxies: starburst
--- galaxies: stellar content --- radiative transfer}

\section{Introduction}

  The intrinsic spectral energy distribution (SED) of a galaxy can
tell us much about that galaxy's stellar evolutionary history, but
dust affects the intrinsic SED in a complicated fashion.  The effects
of dust on a galaxy's SED are determined by the physical properties
and spatial distribution of the dust.  Our knowledge of the physical
properties and spatial distribution of dust in galaxies other than the
Milky Way (MW) and the Magellanic Clouds is very limited.  Starburst
galaxies are good objects in which to study dust as their energetic
output is dominated by massive stars and they are bright in the
ultraviolet (UV) where the effects of dust are large.  In addition,
proper removal of the dust's effects is necessary in order to study
the starburst's stellar evolutionary history (i.e.\ age, star
formation rate, initial mass function, etc.).

  The physical properties of dust affecting a galaxy's SED are
parameterized by the wavelength dependence of the dust's extinction,
albedo, and scattering phase function.  The extinction curve is the
dominant physical property and the easiest to determine
observationally.  Extinction curves for stars in our Galaxy, the Small
\& Large Magellanic Clouds (SMC \& LMC), and (tentatively) M31 have
been determined and found to have a common shape, but exhibit
significant variations.  Generally, the extinction increases with
decreasing wavelength, making the effects of dust strongest in the UV.
The most prominent feature in UV/optical extinction curves is the
2175~\AA\ bump, which varies greatly in strength in our Galaxy
(\cite{wit84}; \cite{car89}) and the LMC (\cite{fit85}), is absent in
the SMC (\cite{pre84}, but see \cite{leq82} for an exception), and
seems weaker in M31 than in our Galaxy (\cite{bia96}).  The behavior
of the 2175~\AA\ bump in the LMC seems particularly relevant to the
question of the type of dust present in starbursts.  The average
extinction curve for the 30~Dor region shows a very weak 2175~\AA\
bump and a strong far-UV rise, while the average extinction curve for
stars outside the 30~Dor region shows a 2175~\AA\ bump strength and
far-UV rise comparable to that of the average Galactic 2175~\AA\ bump
(\cite{cla85}; \cite{fit85}; \cite{fit86}).  This behavior implies the
extinction curve in starbursts may be more like the extinction in the
30~Dor region, lacking a significant 2175~\AA\ bump and having a
strong far-UV rise.  In fact, the 30~Dor region has been described as
the Rosetta Stone for starbursts as it is the most massive star
cluster in the Local Group (\cite{hun95}) and it is close enough to
study individual stars (\cite{wal91}).

  The spatial distribution of the dust plays a crucial role in
determining the strength of dust's effects on the SED of a galaxy.
Witt, Thronson, \& Capuano (1992, hereafter \cite{wit92}) investigated
the effects of different spherical distributions of dust and stars
(termed galactic environments) using a Monte Carlo radiative transfer
model.  Their finding was that the SED of a galaxy will always be
dominated by the least attenuated stars as they contribute the most
flux to the SED.  As a result, the signatures of dust seen in the
galaxy SED will almost always imply small amounts of dust.  Recently,
Witt \& Gordon (1996) studied the radiative transfer in a clumpy
(two-phase) dust distribution.  They found the clumpiness resulted in
more photons escaping than in the equivalent homogeneous case as the
inter-clump medium provides small optical depth paths.  The dust in
galaxies is distributed in clouds (e.g.\ \cite{dic89}; \cite{sca90};
\cite{ros95}), implying clumpiness is necessary to include in any
radiative transfer model involving dust in galaxies.

  In discussing the effects of dust in galaxies, it is important to
distinguish between a galaxy's attenuation curve and the dust's
extinction curve in that galaxy.  The dust's extinction curve is
directly related to the physical properties of the dust grains.  An
extinction curve can be observationally determined by using the
standard pair method, observing both a reddened and unreddened star of
the same spectral type and comparing their fluxes (e.g.\ \cite{mas83};
\cite{wit84}; \cite{fit90}).  The attenuation curve of a galaxy is
defined as the change in the galaxy's flux due to the presence of
dust.  This means the attenuation curve is dependent on both the
physical properties of the dust grains {\em and} the spatial
distribution of the stars and dust.  Thus, the attenuation curve of a
galaxy cannot be used directly to probe the composition of the dust
grains.

  Due to the complexity in studying the dust in the integrated SED of
a galaxy, multiwavelength observations covering a long baseline are
needed.  For the average Milky Way type dust ($R_V = 3.1$), the dust
extinction at 1500~\AA\ is 2.5 times the extinction at 5500~\AA\ (V)
which is 10 times the extinction a 2.2~$\micron$ (K) (\cite{whi92}).
Therefore, multiwavelength observations probe a large dynamical range
of the dust attenuation and allow one to better constrain the dust
spatial distribution and physical properties.

  Multiwavelength observations of starbursts, from the UV to the
K-band, have recently been used to derive the general behavior of dust
attenuation in starburst galaxies (Calzetti, Kinney, \&
Storchi-Bergmann 1994, 1996; \cite{cal97}).  The average attenuation
curve for starbursts was determined using a method similar to the
standard pair method for determining extinction curves.  The starburst
SEDs were collected into bins according to their ionized gas
$E(B-V)_i$ values (\cite{sto94}; \cite{cal94}) and attenuation curves
were determined using the smallest $E(B-V)_i$ bin as the unattenuated
stellar population.  When scaled by their respective $E(B-V)_i$
values, the attenuation curves were roughly similar in shape and
strength, and this is the behavior of dust in a screen-like geometry.
When normalized to an E(B-V)~=~1, the resulting average starburst
attenuation curve has a slope similar to the MW extinction curve, but
lacks a 2175~\AA\ bump, like the SMC extinction curve.  Generally, the
dust distribution can be described by a distribution of foreground
clumps.  The strength of the attenuation affecting the stars was found
to be different than that affecting the gas, implying that the stars
have a dust covering factor smaller (by $\sim$40\%) than the gas.
This implies the stars and gas have different spatial distributions.
While the flatter attenuation curve can be explained as an effect of
the radiative transfer in the dust, we will argue in this paper the
lack of a 2175~\AA\ bump in the average attenuation curve can only be
explained by a dust extinction curve with a weak or missing bump.

  Without prior knowledge of the dust's type and distribution or
stellar evolutionary history in starbursts, the only way to
investigate the SED of a starburst is to combine a stellar
evolutionary synthesis model with a model for the radiative transfer
in dust and simultaneously solve for the characteristics of the stars
and dust.  The aim of this work is to investigate a large sample of
starbursts and derive the geometry and type of dust present in
starbursts.  In future work, this will allow us to investigate with
greater confidence the evolutionary history of individual starbursts.
In $\S$\ref{sec_model}, we discuss our model of a starburst galaxy
which includes stellar evolutionary synthesis and the radiative
transfer through the starburst's dust.  The observational sample
(consisting of 30 starburst galaxy SEDs) used in this paper is
described in $\S$\ref{sec_obs}.  Section~\ref{sec_cc_plots} presents
color-color plots useful for detecting dust in starbursts as well as
constraining the type and spatial distribution of the dust.  As an
example, we apply our starburst model to the galaxies NGC~4385 \&
NGC~7714 in $\S$\ref{sec_ngc7714}.  We discuss our results and their
implications in $\S$\ref{sec_discussion}.  Finally, our conclusions
are presented in $\S$\ref{sec_conclusions}.

\section{Starburst Model \label{sec_model}}

  In order to model the SED of starburst galaxies, both a stellar
evolutionary synthesis model and a dust radiative transfer model are
needed.  We combine the stellar evolutionary synthesis models of
Leitherer \& Heckman (1995; hereafter \cite{lei95}) and Bruzual \&
Charlot (1993, 1997; hereafter \cite{bru97}) to obtain the intrinsic
SEDs of both continuous and instantaneous bursts of star formation
with ages between $1 \times 10^6$ and $15 \times 10^{9}$~years.  We
use a radiative transfer code based on Monte Carlo techniques to model
the effects of dust on the intrinsic SEDs for both MW and SMC type
dust.

\subsection{Radiative Transfer in Dust}

   In order to accurately model the radiative transfer of the stellar
photons through the dust, we rely on Monte Carlo techniques
(\cite{wit77}).  With these techniques, photons are followed through a
dust distribution and their interaction with the dust is parameterized
by the dust optical depth ($\tau$), albedo, and scattering phase
function asymmetry (g).  The optical depth determines where the photon
interacts, the albedo gives the probability the photon is scattered
from a dust grain, and the scattering phase function gives the angle
at which the photon scatters.  As the physical properties of dust in
different environments vary greatly (\cite{wit84}; \cite{car89};
\cite{fit89}; \cite{mat92}), we have chosen to model two extremely
different kinds of dust: MW and SMC.  For both kinds of dust, we have
compiled the wavelength dependence (1200--23,000~\AA) of the optical
depth, albedo, and g in Table~\ref{table_dust_phys}.  For MW type
dust, the $\tau/\tau_V$ values are from Whittet (1992) and the albedo
and g values were determined from smooth curves drawn through many
empirical determinations taken from the literature (e.g.\ \cite{gor94}
[and references therein]; \cite{cal95}; \cite{leh96}).  For SMC type
dust, the $\tau/\tau_V$ values are from Pr\'evot et al.\ (1984) and
the albedo and g values were kindly computed for us by Sang-Hee Kim
(\cite{kim94}; \cite{kim96}) assuming that the SMC extinction curve is
produced by a mixture of silicate and graphite grains.  The
characteristics of the magnitude bands (UV1 to K) are given in
Table~\ref{table_mag_def}.

\begin{deluxetable}{clccccc}
\tablewidth{0pt}
\tablecaption{Dust Physical Properties \label{table_dust_phys}}
\tablehead{\colhead{} &
           \multicolumn{3}{c}{MW} & \multicolumn{3}{c}{SMC} \\
           \colhead{band} & 
           \colhead{$\tau/\tau_V$} & \colhead{albedo} & \colhead{g} & 
           \colhead{$\tau/\tau_V$} & \colhead{albedo} & \colhead{g}}
\startdata
UV1 & 3.11  & 0.60 & 0.75 & 5.00 & 0.40 & 0.53 \nl
UV2 & 2.63  & 0.67 & 0.75 & 4.36 & 0.40 & 0.53 \nl
UV3 & 2.50  & 0.65 & 0.73 & 3.51 & 0.58 & 0.54 \nl
UV4 & 2.78  & 0.55 & 0.72 & 3.20 & 0.58 & 0.51 \nl
UV5 & 3.12  & 0.46 & 0.71 & 2.90 & 0.55 & 0.46 \nl
UV6 & 2.35  & 0.56 & 0.70 & 2.40 & 0.56 & 0.37 \nl
UV7 & 2.00  & 0.61 & 0.69 & 2.13 & 0.53 & 0.35 \nl
U   & 1.52  & 0.63 & 0.65 & 1.58 & 0.46 & 0.34 \nl
B   & 1.32  & 0.61 & 0.63 & 1.35 & 0.43 & 0.32 \nl
V   & 1.00  & 0.59 & 0.61 & 1.00 & 0.43 & 0.29 \nl
R   & 0.76  & 0.57 & 0.57 & 0.74 & 0.41 & 0.26 \nl
I   & 0.48  & 0.55 & 0.53 & 0.52 & 0.38 & 0.23 \nl
J   & 0.28  & 0.53 & 0.47 & 0.28 & 0.33 & 0.21 \nl
H   & 0.167 & 0.51 & 0.45 & 0.17 & 0.30 & 0.23 \nl
K   & 0.095 & 0.50 & 0.43 & 0.11 & 0.29 & 0.22 \nl
\enddata
\end{deluxetable}

  The spatial distribution of the stars and dust was investigated by
assuming either dusty or shell geometries.  We expect these geometries
to bracket the large range of observational star/dust distributions.
Following \cite{wit92}, the dusty geometry has stars and dust
uniformly mixed through the entire sphere and the shell geometry
(nuclei geometry in \cite{wit92}) has a dust-free sphere of stars ($r
< 0.3r_{\rm max}$) surrounded by a star-free shell of dust ($0.3r_{\rm
max} < r < r_{\rm max}$).  Unlike \cite{wit92}, we also allow the dust
to be either homogeneous or clumpy (two-phases) on a local scale (but
see \cite{wit97} for an extension of \cite{wit92} to include clumpy
dust).  The two-phase clumpy dust is characterized by the ratio of
inter-clump to clump medium density ($k_2/k_1$), the filling factor
(ff) of the high density clumps, and the size of the clumps compared
to the system size (\cite{wit96}).

  The radiative transfer model has the following inputs: the geometry
(dusty or shell), homogeneous or clumpy dust distribution ($k_2/k_1$,
ff, \& clump size), radial $\tau$ in the V band (equivalent to dust
mass), and either MW or SMC type dust grain properties ($\tau$,
albedo, and g [see Table~\ref{table_dust_phys}]).  The different model
runs are listed in Table~\ref{table_mod_runs}.  All model runs were
computed for ff $= 0.15$ (\cite{wit96}), a clump size of 10\% of the
system size (radius), and $0.05 \leq \tau_V \leq 40$.

\begin{deluxetable}{cccc}
\tablewidth{0pt}
\tablecaption{Model Run Inputs \label{table_mod_runs}}
\tablehead{\colhead{run} & \colhead{geometry} & \colhead{$k_2/k_1$} & 
           \colhead{dust type}}
\startdata
1 & shell & 1.0 & MW \nl
2 & shell & 0.1 & MW \nl
3 & shell & 0.01 & MW \nl
4 & shell & 0.001 & MW \nl
5 & shell & 1.0 & SMC \nl
6 & shell & 0.1 & SMC \nl
7 & shell & 0.01 & SMC \nl
8 & shell & 0.001 & SMC \nl
9 & dusty & 1.0 & MW \nl
10 & dusty & 0.1 & MW \nl
11 & dusty & 0.01 & MW \nl
12 & dusty & 0.001 & MW \nl
13 & dusty & 1.0 & SMC \nl
14 & dusty & 0.1 & SMC \nl
15 & dusty & 0.01 & SMC \nl
16 & dusty & 0.001 & SMC \nl
\enddata
\end{deluxetable}

\subsection{Stellar Evolutionary Synthesis \label{sec_star_model}}

  The description of the stellar evolution of the starbursts is
derived from the work of \cite{lei95} and \cite{bru97}.

  \cite{lei95} concentrated on the stellar evolutionary synthesis of
massive stars, including nebular emission associated with massive
stars.  They produce SEDs for the two different extreme cases of star
formation, instantaneous and constant.  For both cases of star
formation, we use an initial mass function (IMF) with a \cite{sal55}
slope ($\alpha = 2.35$), a mass range of 1--100~M$_{\sun}$, and solar
metallicity.  We have adjusted the absolute level of the SEDs to
reflect a mass range of 0.1--100~M$_{\sun}$ using equation 3 in
\cite{lei95}.  Since their model lacks information about the later
stages of stellar evolution, we limit our use of their SEDs to those
with ages less than $2 \times 10^7$~years.

  \cite{bru97} present a stellar evolutionary synthesis code which
produces SEDs for both instantaneous and constant star formation.  The
SEDs are formed using an IMF with a \cite{sal55} slope ($\alpha =
2.35$), a mass range of 0.1--125~M$_{\sun}$, and solar metallicity.
As this code lacks information on the nebular continuum, we limit our
use of these SEDs to the ages between $5 \times 10^7$ and $1.5 \times
10^{10}$~years.  In the continuous star formation case, where the
nebular contribution from the most recent generation of stars is
present, we add the nebular continuum calculated from a \cite{lei95}
model of constant star formation for an age of $2 \times 10^7$~years.

  While both \cite{lei95} and \cite{bru97} provide SEDs computed for
metallicities other than solar, we have chosen to use only the solar
metallicity SEDs for three reasons.  First, the average metallicity of
the observed starbursts is relatively high, around 0.5 solar (see
$\S$\ref{sec_obs}).  Second, theoretical models of low metallicity
stellar populations still have difficulties in reproducing all the
observational constraints (e.g., the ratio of blue/red supergiants,
see Maeder \& Conti [1994]).  Third, using model SEDs with a higher
metallicity than actually observed will result in a lower limit on the
effects of dust.  This is due to the fact that the lower the
metallicity the bluer the SED (\cite{wor94}) and since effects of dust
are to redden the SED, using solar metallicity SEDs will give a lower
limit on the dust's effects.  In future work, we will use SEDs of the
observed metallicities when investigating the stellar evolutionary
history of individual starbursts.

   The upper mass limit on the IMF between the two models is
different; 100~M$_{\sun}$ for \cite{lei95} and 125~M$_{\sun}$ for
\cite{bru97}.  This is not expected to be important as the IMF is
steep, the amount of mass between 100 and 125~$M_{\sun}$ is less than
1\% of the total mass, and its relative contribution to the
ionization and heating of the interstellar medium is small.

  Using the above stellar evolutionary models, SEDs from
1000--23,000~\AA\ were generated with ages between $1 \times 10^6$ and
$15 \times 10^{9}$~years.  These SEDs were convolved with photometric
bandpasses at UV, optical, and IR wavelengths to generate the
intrinsic colors of the starburst model galaxies.  In the UV, 7 square
bandpasses (UV1--UV7) were defined with standard UV zero magnitude
fluxes (\cite{wes82}).  For the optical and IR, photometric bandpasses
for UBVRIJHK were taken from \cite{bes88} and Bessell (1990).  The
zero magnitude fluxes for the UBVRI bands were calculated by
convolving the appropriate bandpass with the observed spectrum of Vega
(\cite{tug77}).  The calibrated spectrum of Vega was multiplied by
1.028 before use to account for the fact that Vega's V magnitude is
0.03 (\cite{hof91}).  The zero magnitude fluxes for the JHK bands were
taken from (\cite{bes90}).  The $\lambda_{\rm eq}$, $\Delta\lambda$,
and zero magnitude fluxes for the adopted bandpasses are presented in
Table~\ref{table_mag_def}.

\begin{deluxetable}{crrc}
\tablewidth{0pt}
\tablecaption{Magnitude Band Definitions \label{table_mag_def}}
\tablehead{\colhead{band} & \multicolumn{1}{c}{$\lambda_{\rm eq}$} & 
           \multicolumn{1}{c}{$\Delta\lambda$} & 
           \colhead{F$_{\lambda}$\tablenotemark{a}} \\
           \colhead{} & \colhead{[\AA]} & \colhead{[\AA]} & 
           \colhead{[ergs cm$^{-2}$ s$^{-1}$ \AA$^{-1}$]}}
\startdata
UV1 & 1250 &  250 & 3.63\sn{-9} \nl
UV2 & 1515 &  280 & 3.63\sn{-9} \nl
UV3 & 1775 &  260 & 3.63\sn{-9} \nl
UV4 & 1995 &  200 & 3.63\sn{-9} \nl
UV5 & 2215 &  260 & 3.63\sn{-9} \nl
UV6 & 2480 &  290 & 3.63\sn{-9} \nl
UV7 & 2895 &  560 & 3.63\sn{-9} \nl
U   & 3605 &  640 & 3.41\sn{-9} \nl
B   & 4413 &  959 & 6.60\sn{-9} \nl
V   & 5512 &  893 & 3.70\sn{-9} \nl
R   & 6594 & 1591 & 2.35\sn{-9} \nl
I   & 8059 & 1495 & 1.15\sn{-9} \nl
J  & 12369 & 2034 & 3.12\sn{-10} \nl
H  & 16464 & 2862 & 1.14\sn{-10} \nl
K  & 21578 & 2673 & 3.94\sn{-11} \nl
\enddata
\tablenotetext{a}{F$_{\lambda}$ is the flux corresponding to a
magnitude of zero.  See text for details.}
\end{deluxetable}

\section{Observations \label{sec_obs}}

  For this paper, we use a sample of 30 starburst galaxies which have
SEDs available in the literature.  For consistency, we provide a brief
summary of the characteristics of galaxies and the properties of the
SEDs.  We refer the reader to the original papers for the
observational details (\cite{kin93}; \cite{mcq95}; \cite{sto95};
\cite{cal97}).

  The galaxies are from the {\em IUE} Atlas of Kinney et al.\ (1993)
and are characterized by the presence of active star formation in
their central regions. The star formation activity manifests itself as
strong UV SEDs and intense line emission at optical and IR
wavelengths.  The galaxies are nearby (median distance $60$~Mpc for
H$_o$ = 50~km~s$^{-1}$~Mpc$^{-1}$) irregulars or spirals with
disturbed morphologies.  The metallicity of the starbursts' ionized
gas is in the range 0.1--2~solar, with a median around 0.5~solar.

  Aperture-matched UV and optical SEDs (\cite{kin93}; \cite{mcq95};
\cite{sto95}) are available for all 30 galaxies and aperture-matched
IR photometry at J, H, and K (\cite{cal97}) is available for 19 of the
galaxies.  The {\em IUE} aquired UV portion of the SED was corrected
to reflect the newest {\em IUE} absolute calibration (\cite{boh96}).
The match in observational apertures ensured that the same region was
sampled within each galaxy.  Since the UV SEDs were acquired with {\em
IUE}, apertures of $10\arcsec \times 20\arcsec$ were used for the
optical and IR observations.  Differences in orientation between the
IUE and optical apertures produced an estimated 10--20\% flux mismatch
between the two wavelength regions.  This size aperture subtends large
regions within each galaxy, a diameter of $\sim$4.5~kpc at the median
distance of 60~Mpc.

  In most cases, the UV and optical SEDs cover the wavelength range
1250--7500~\AA\ continuously with a resolution 6~\AA\ in the UV and
$\sim$10~\AA\ in the optical. The availability of spectral information
allowed us to assess the importance of the emission line contribution
to the broad band photometry. The IR images (\cite{cal97}) were used
to obtain the J, H, and K photometry with in an area of $10\arcsec
\times 20\arcsec$ in the E--W direction.  In the IR, the contribution
of the emission lines to the broad band photometry is negligible,
except for the J-band photometry of NGC~4861, where the hydrogen
recombination line Pa$\beta$ (12818~\AA) contributes 9\% of the band
flux (\cite{cal97}).

  As with the SEDs produced with the stellar evolutionary models, the
SEDs of the 30 starbursts were convolved with photometric bandpasses
at UV and optical wavelengths to determine their UV2--I photometry.
Prior to this, the SEDs were corrected for foreground Galactic
extinction, shifted to zero redshift, and cleaned of emission lines.
The dereddening was done using the average Galactic extinction curve
(\cite{whi92}) and the values of $E(B-V)_G$ listed in Table~1 of
Calzetti, Kinney, \& Storchi-Bergmann (1994).  The SEDs were shifted
to zero redshift using the values of $z$ listed in the same table.
The emission lines were removed as the stellar evolutionary synthesis
model SEDs did not include emission lines and we wish to compare the
observations to these models.  The IR photometry was transformed from
the CIT system (\cite{eli82}) to the homogeneous system defined by the
equations in Bessel \& Brett (1988).  The resulting magnitudes from
the UV, optical, and IR are tabulated in Table~\ref{table_photometry}.
The uncertainties in the tabulated magnitudes are 0.10 for UV2--UV7
(\cite{boh90}; \cite{boh97}), 0.10--0.20 for U--I (0.10 for those
galaxies with both UV7 and U data), and 0.09--0.14 for J--K
(\cite{kin93}; \cite{mcq95}; \cite{sto95}; \cite{cal97}).  Due to
mismatches between the apertures used in obtaining the optical and IR
photometry, the relative uncertainties between these two wavelength
regions are 0.10--0.20 magnitudes.  In all the plots in this paper, we
have assumed a uncertainty of 0.10 for UV2--UV7 data, 0.10 for U--I
for galaxies with UV7 and U data, 0.15 for U--I for galaxies without
both UV7 and U, and 0.125 for J--K.  For colors involving only J--K,
we have used a uncertainty of 0.07 for (J-H) and 0.05 for (H-K)
(\cite{cal97}).

\begin{deluxetable}{lcccccccccccccc}
\scriptsize
\tablewidth{0pt}
\tablecaption{Starburst Galaxy Photometry \label{table_photometry}}
\tablehead{\colhead{name} & \colhead{UV2} & \colhead{UV3} & \colhead{UV4} & 
    \colhead{UV5} & \colhead{UV6} & \colhead{UV7} & \colhead{U} & 
    \colhead{B} & \colhead{V} & \colhead{R} & \colhead{I} & 
    \colhead{J} & \colhead{H} & \colhead{K}}
\startdata
IC 1586 & 14.27 & 14.34 & \nodata & \nodata & \nodata & \nodata & 14.77 & 15.32 & 14.92 & \nodata & \nodata & 13.58 & 12.83 & 12.52 \nl
HARO 15 & 12.99 & 13.14 & \nodata & \nodata & \nodata & \nodata & 14.37 & 15.02 & 14.78 & 14.53 & 14.05 & 13.54 & 12.92 & 12.58 \nl
MRK 357 & 13.29 & 13.55 & \nodata & \nodata & \nodata & \nodata & 14.90 & 15.87 & 15.79 & \nodata & \nodata & 14.55 & 13.86 & 13.39 \nl
IC 214 & 14.10 & 14.14 & \nodata & \nodata & \nodata & \nodata & 15.59 & 16.03 & 15.48 & \nodata & \nodata & 12.66 & 11.85 & 11.37 \nl
NGC 1140 & 11.96 & 12.17 & \nodata & \nodata & \nodata & \nodata & 13.11 & 13.75 & 13.43 & 13.18 & 12.74 & 12.54 & 11.86 & 11.62 \nl
NGC 1510 & 12.91 & 13.14 & 13.42 & 13.56 & 13.87 & 14.02 & 14.33 & 14.87 & 14.55 & 14.25 & \nodata & \nodata & \nodata & \nodata \nl
NGC 1569 & 9.68 & 9.72 & 9.95 & 10.18 & 10.19 & 10.46 & 10.71 & 11.66 & 11.47 & 11.36 & \nodata & 10.93 & 10.29 & 10.12 \nl
NGC 1614 & 14.16 & 14.16 & \nodata & \nodata & \nodata & \nodata & 14.18 & 14.57 & 13.91 & 13.36 & 12.57 & 11.46 & 10.60 & 10.07 \nl
NGC 1705 & 10.55 & 10.94 & 11.23 & 11.40 & 11.82 & 12.27 & 12.70 & 13.61 & 13.43 & 13.22 & 12.79 & \nodata & \nodata & \nodata \nl
NGC 3049 & 13.54 & 13.58 & 13.80 & 13.92 & 14.08 & 14.22 & 14.44 & 15.09 & 14.64 & 14.30 & 13.82 & \nodata & \nodata & \nodata \nl
NGC 3125 & 12.24 & 12.33 & 12.68 & 12.85 & 13.10 & 13.34 & 13.50 & 14.32 & 14.08 & 13.73 & 13.24 & \nodata & \nodata & \nodata \nl
NGC 4194 & 13.39 & 13.31 & 13.44 & 13.52 & 13.57 & 13.48 & 13.40 & 13.96 & 13.35 & \nodata & \nodata & 11.43 & 10.67 & 10.24 \nl
NGC 4385 & 13.30 & 13.29 & 13.55 & 13.53 & 13.87 & 13.98 & 14.17 & 14.71 & 14.09 & 13.47 & 12.82 & 12.14 & 11.44 & 11.12 \nl
NGC 4861  & 11.43 & 11.78 & 12.17 & 12.54 & 12.77 & 13.07 & \nodata & \nodata & 14.91 & 14.89 & \nodata & 13.83 & 13.41 & 13.10 \nl
NGC 5236 & 10.32 & 10.30 & 10.43 & 10.48 & 10.70 & 10.83 & 11.02 & 11.60 & 11.13 & 10.80 & 10.23 & \nodata & \nodata & \nodata \nl
NGC 5253 & 10.61 & 10.75 & 10.91 & 11.07 & 11.35 & 11.65 & 12.05 & 12.93 & 12.75 & 12.53 & 12.24 & \nodata & \nodata & \nodata \nl
UGC 9560 & 12.69 & 12.98 & 13.32 & 13.54 & 13.76 & 14.13 & 14.65 & 15.59 & 15.32 & 15.13 & \nodata & \nodata & \nodata & \nodata \nl
NGC 5860 & 14.23 & 14.19 & 14.71 & 14.76 & 14.79 & 14.66 & 14.57 & 14.91 & 14.34 & \nodata & \nodata & 12.87 & 12.12 & 11.83 \nl
NGC 5996 & 13.36 & 13.37 & 13.67 & 13.56 & 13.84 & 13.97 & 14.10 & 14.74 & 14.28 & \nodata & \nodata & \nodata & \nodata & \nodata \nl
NGC 6052 & 13.36 & 13.36 & 13.59 & 13.72 & 13.84 & 13.96 & 13.81 & 14.42 & 14.03 & 13.91 & \nodata & 12.77 & 12.13 & 11.82 \nl
NGC 6090 & 13.86 & 13.82 & \nodata & \nodata & \nodata & \nodata & 14.58 & 15.14 & 14.64 & 14.31 & \nodata & 12.65 & 11.92 & 11.46 \nl
NGC 6217 & 13.04 & 12.93 & \nodata & \nodata & \nodata & \nodata & \nodata & 14.11 & 13.55 & 13.15 & \nodata & 11.55 & 10.80 & 10.45 \nl
TOL 1924-416 & 11.90 & 12.19 & 12.50 & 12.73 & 13.03 & 13.33 & 13.49 & 14.27 & 13.95 & 13.74 & 13.45 & \nodata & \nodata & \nodata \nl
NGC 7250 & 12.42 & 12.51 & 12.80 & 12.69 & 13.06 & \nodata & 13.69 & 14.48 & 14.17 & 13.94 & \nodata & 13.00 & 12.39 & 12.12 \nl
NGC 7552 & 13.34 & 13.06 & 13.12 & 13.03 & 13.02 & 12.94 & 12.65 & 12.99 & 12.38 & 11.90 & 11.17 & \nodata & \nodata & \nodata \nl
NGC 7673 & 12.72 & 12.85 & 13.12 & 13.31 & 13.53 & 13.69 & 13.61 & 14.19 & 13.82 & 13.60 & \nodata & 12.63 & 12.02 & 11.73 \nl
NGC 7714 & 12.18 & 12.27 & 12.56 & 12.71 & 12.84 & 12.95 & 13.11 & 13.81 & 13.43 & 13.07 & 12.59 & 11.87 & 11.17 & 10.81 \nl
NGC 7793 & 13.64 & 13.77 & 14.02 & 14.27 & 14.44 & 14.40 & 14.22 & 14.60 & 13.99 & 13.58 & 12.97 & \nodata & \nodata & \nodata \nl
MRK 309 & 14.92 & 14.79 & \nodata & \nodata & \nodata & \nodata & 15.13 & 15.81 & 15.43 & \nodata & \nodata & 13.77 & 13.05 & 12.54 \nl
MRK 542 & 14.21 & 14.37 & \nodata & \nodata & \nodata & \nodata & 15.51 & 15.80 & 15.27 & \nodata & \nodata & 13.66 & 13.01 & 12.66 \nl
\enddata
\end{deluxetable}

\section{Color-Color Plots \label{sec_cc_plots}}

  The main problem with accurately modeling the ages of the stellar
populations and effects of dust in a galaxy is that aging the stellar
population and adding dust both redden the galaxy's SED.  A
non-exhaustive search of the {\it many} color-color plots turned up a
handful of plots where the effects of stellar population changes were
in a different direction in the color-color plane than those due to
the dust.  By examining where the sample of observed starburst
galaxies fall on these color-color plots, information about the dust
(type/geometry) and the stellar population (burst/constant and age)
can be derived.

\begin{figure}[tbp]
\begin{center}
\plotone{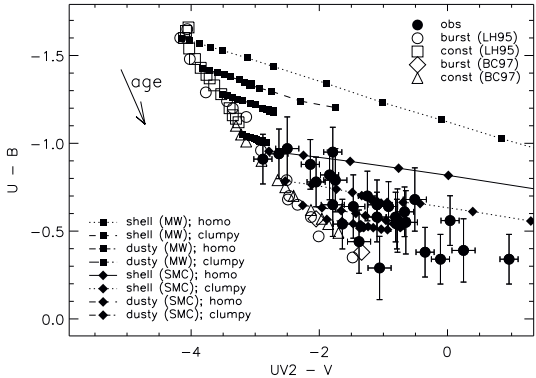}
\caption{The color-color plane (U-B) vs.\ (UV2-V) is plotted for both
the starburst observations and models.  The dust reddening
trajectories can start from any stellar evolutionary synthesis model,
but to plot them, they were attached to an arbitrary stellar
evolutionary synthesis model.  The clumpy geometry refers to a
$k_2/k_1 = 0.01$.  The words burst and const in the upper legend refer
to the kind of star formation (instantaneous or constant,
respectively) assumed in the stellar evolutionary synthesis
models. The age vector points in the direction of increasing age of
the stellar evolutionary synthesis SEDs.
\label{fig_ub_uv2v}}
\end{center}
\end{figure}

  The color-color plot (U-B) vs.\ (UV2-V) is plotted in
Figure~\ref{fig_ub_uv2v} and plainly shows the presence of dust in
starburst galaxies.  All of the stellar evolutionary synthesis models
fall in a well defined line, with age increasing downward, while the
actual observations all lie to the right of this line.  Combining
different stellar evolutionary models to approximate a starburst with
a complex stellar evolutionary history will not move the starburst off
of the line defined by the simple stellar evolutionary models.  All of
the reddening trajectories point to the right but some are not long
enough to explain the spread in the observed starbursts.  Those which
cannot explain the spread are the dusty geometry models which assume
the same global distributions of stars and dust, be they homogeneous or
clumpy.  The shell geometry can explain the observed spread and we
adopt this as the geometry for starburst galaxies.  This geometry
results in a screen-like behavior of the dust which allowed Calzetti
et al.\ (1994) to derive an average attenuation curve for starbursts
in a manner analogous to the standard pair method.  It is important to
remember that an attenuation curve is an extinction curve convolved
with the dust geometry and thus an attenuation curve cannot be used to
probe the type of dust grains directly.

  From Figure~\ref{fig_ub_uv2v}, the stellar population can be
constrained by moving along the reddening trajectories back to the
line defined by the stellar evolutionary models.  The stellar
populations seem to arise from burst star formation models with ages
between $8 \times 10^{6}$ and $2 \times 10^{8}$~years or constant star
formation models between $1 \times 10^{8}$ and $15 \times
10^{9}$~years.  Actually determining the star formation history of an
individual galaxy is quite complex and requires fitting the SED from
the UV to the IR.  See $\S$\ref{sec_ngc7714} for an example of how
this can be done.  The absence of starbursts (from our UV selected
sample) with ages $<~8 \times 10^{6}$~years suggests the initial phase
of starbursts is hidden within dust clouds and is difficult to observe
in the UV.  In addition, the age we derive is the average age of the
starburst regions falling in the large aperture used to obtain the
SEDs.  Thus, starburst regions with ages $<~8 \times 10^{6}$~years
could be present, but washed out by the presence of older starburst
regions.

\begin{figure}[tbp]
\begin{center}
\plotone{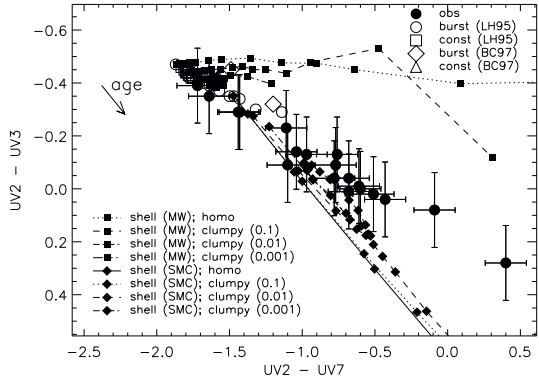}
\caption{The color-color plane (UV2-UV3) vs.\ (UV2-UV7) is plotted.
The number in parenthesis after the word clumpy in the lower legend
refers to the value of $k_2/k_1$.  Otherwise, this figure was
constructed in a manner similar to Figure~1.
\label{fig_uv2uv3_uv2uv7}}
\end{center}
\end{figure}

\begin{figure}[tbp]
\begin{center}
\plotone{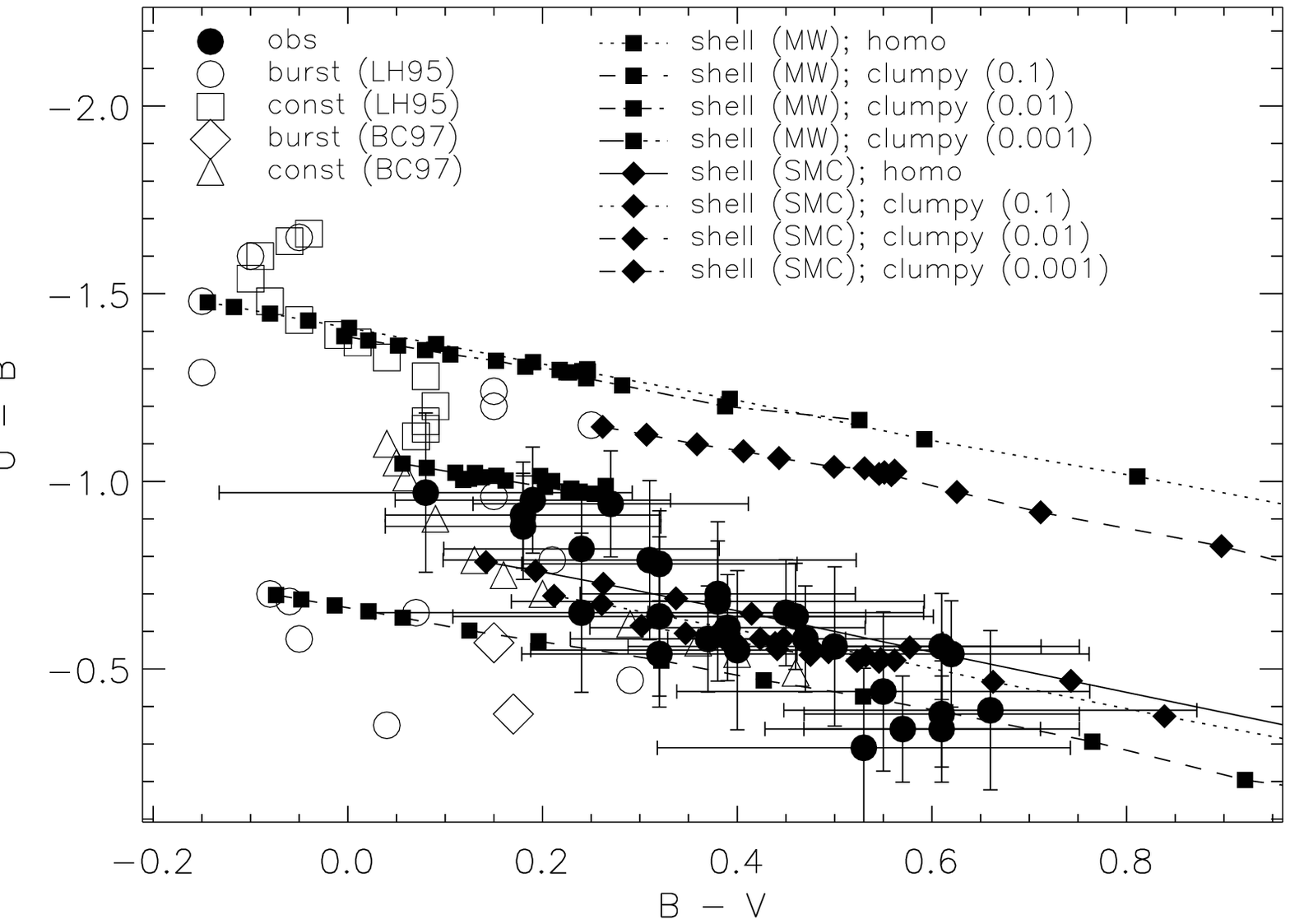}
\caption{The color-color plane (U-B) vs.\ (B-V) is plotted. The number
in parenthesis after the word clumpy in the right legend refers to the
value of $k_2/k_1$.  Otherwise, this figure was constructed in a
manner similar to Figure~1.  \label{fig_ub_bv}}
\end{center}
\end{figure}

\begin{figure}[tbp]
\begin{center}
\plotone{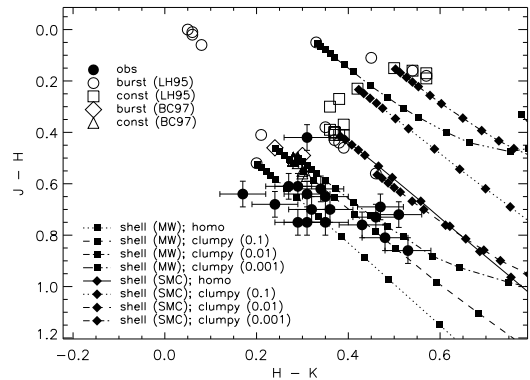}
\caption{The color-color plane (J-H) vs.\ (H-K) is plotted. The number
in parenthesis after the word clumpy in the lower legend refers to the
value of $k_2/k_1$.  Otherwise, this figure was constructed in a
manner similar to Figure~1.  \label{fig_jh_hk}}
\end{center}
\end{figure}

  Figures~\ref{fig_uv2uv3_uv2uv7}--\ref{fig_jh_hk} show examples of
useful color-color plots (one each for the UV, optical, and IR) which
show differences between stellar population changes and dust
reddening.  The (UV2-UV3) vs.\ (UV2-UV7) color-color plane
(Figure~\ref{fig_uv2uv3_uv2uv7}) is a good plane in which to
investigate whether SMC or MW type dust best matches the type of dust
in starbursts.  The observations fall between the SMC and MW reddening
trajectories, but closest to the SMC reddening trajectories.  This
implies the far-UV slope of the starbursts' extinction curve has a
value between the MW and SMC, but closest to the SMC.  In
$\S$\ref{sec_2175}, we investigate the starbursts' UV extinction curve
in more detail, focusing on the 2175~\AA\ bump.  The optical (U-B)
vs.\ (B-V) color-color plane (Figure~\ref{fig_ub_bv}) does show that
dust is present, but the effects are not as definitive as those shown
in the previous two figures.  This points to the need, in investigating
the dust in starbursts, for SEDs covering more than just the optical.
The (J-H) vs.\ (H-K) color-color plane (Figure~\ref{fig_jh_hk}) also
shows the effects of dust, but some degeneracy exists between changes
in age and the effects of dust.  In addition, not all of the
starbursts can be traced back to a stellar population model through
the reddening trajectories.  This is probably due to problems with
stellar evolutionary models in the IR region (see \cite{lei95} and
Lan\c{c}on \& Rocca-Volmerange [1996] for details).  While all of the
above color-color plots (Figures~\ref{fig_ub_uv2v}--\ref{fig_jh_hk})
clearly show the presence of dust in starbursts, the diagnostic
utility of color-color plots is greatest in the UV.

\subsection{Absence of the 2175 \AA\ bump \label{sec_2175}}

\begin{figure}[tbp]
\begin{center}
\plotone{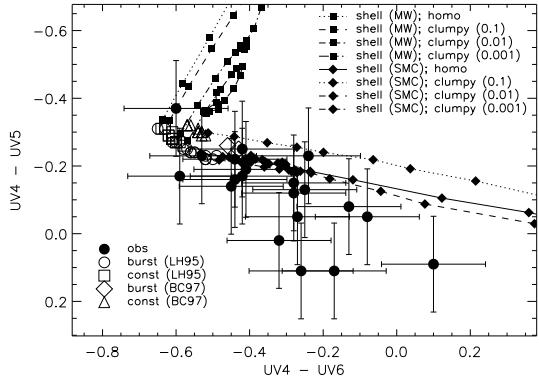}
\caption{The color-color plane (UV4-UV5) vs.\ (UV4-UV6) is
plotted. The number in parenthesis after the word clumpy in the lower
legend refers to the value of $k_2/k_1$.  Otherwise, this figure was
constructed in a manner similar to Figure~1.
\label{fig_uv4uv5_uv4uv6}}
\end{center}
\end{figure}

\begin{figure}[tbp]
\begin{center}
\plotone{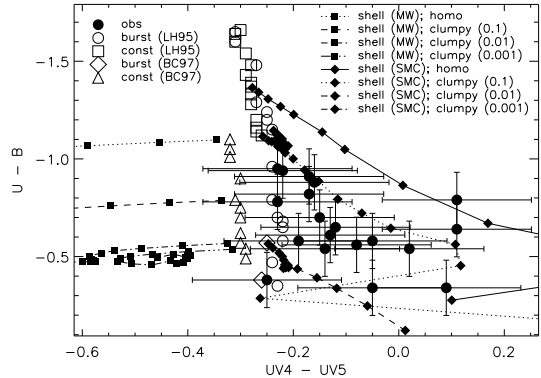}
\caption{The color-color plane (U-B) vs.\ (UV4-UV5) is plotted. The
number in parenthesis after the word clumpy in the lower legend refers
to the value of $k_2/k_1$.  Otherwise, this figure was constructed in
a manner similar to Figure~1.  \label{fig_ub_uv4uv5}}
\end{center}
\end{figure}

  Due to its prominence, the 2175~\AA\ extinction bump is an excellent
probe of the type of dust in a galaxy.  This is illustrated by the
average extinction curves of the MW, LMC, SMC, and M31, all of which
show different 2175~\AA\ strengths (\cite{fit89}; \cite{bia96}).  The
lack of a depression at 2175~\AA\ in SEDs of starbursts has raised the
question: does the dust in starbursts lack a 2175~\AA\ bump or are
there radiative transfer effects associated with the spatial
distribution of the dust which weaken the bump below observational
detection?  To investigate this question, two color-color plots were
constructed.  Figures~\ref{fig_uv4uv5_uv4uv6} \& \ref{fig_ub_uv4uv5}
show the (UV4-UV5) vs.\ (UV4-UV6) and (U-B) vs.\ (UV4-UV5) color
planes, respectively.  The placement of the three bandpasses of UV4,
UV5, \& UV6 ($\lambda_{\rm eq}$ = 1995~\AA, 2215~\AA, \& 2480~\AA)
make them ideal for studying the 2175~\AA\ bump.  The color (UV4-UV6)
measures the slope underneath the 2175~\AA\ bump and the color
(UV4-UV5) is sensitive to the short wavelength slope of the 2175~\AA\
bump.

   Both figures clearly show the SMC reddening trajectories generally
point toward the starburst observations, while the MW reddening
trajectories completely miss the observations.  In
Figure~\ref{fig_uv4uv5_uv4uv6} the MW reddening trajectories point
almost $90\arcdeg$ from the observations and in
Figure~\ref{fig_ub_uv4uv5} they point almost $180\arcdeg$ from the
observed starbursts.  One of the MW reddening trajectories shows how
radiative transfer effects can reduce the strength of the 2175 \AA\
bump.  In Figure~\ref{fig_uv4uv5_uv4uv6}, the reddening trajectory for
model run 4 (shell, MW clumpy with $k_2/k_1 = 0.001$) loops back
toward the stellar evolutionary models.  This is caused by the output
spectrum being dominated by unattenuated stars which suppress both the
2175~\AA\ bump as well as the slope of the attenuation curve in the
UV.  While the starbursts do not show the presence of the 2175~\AA\
bump, they do show the signature of a substantial UV slope in their
attenuation curves.  Therefore, radiative transfer effects alone
cannot explain the lack of a 2175~\AA\ bump in starbursts and, so, the
dust in the average starburst is more like the dust in the SMC than in
the MW.  This agrees with the notion that the dust in starbursts would
be similar to the weak 2175~\AA\ bump 30~Dor region dust, the
mini-starburst next door (\cite{wal91}).

  In Figure~\ref{fig_uv4uv5_uv4uv6}, there are a number of starbursts
which cannot be traced back to the stellar evolutionary models through
a reddening trajectory.  This is not surprising as we have only
investigated two extinction curves and the extinction curves of dust
in the MW, LMC, and SMC are known to possess large variations
(\cite{leq82}; \cite{pre84}; \cite{wit84}; \cite{fit85};
\cite{car89}).  In order to fit these outlying starbursts, we would
need an extinction curve in which the color excess in (UV4-UV5) is
greater for a given color excess in (UV4-UV6) than that in the SMC
extinction curve.  This implies an extinction curve with a curvature
between UV4 and UV6 greater than curvature in the same region in the
SMC extinction curve.  Confirmation of this speculation awaits future
work where we plan to individually fit each of the starbursts and
derive the average starburst extinction curve.

\section{NGC~4385 \& NGC~7714 \label{sec_ngc7714}}

  Fitting an individual galaxy's SED is difficult as there are usually
many starburst models (combinations of stellar evolutionary synthesis
and radiative transfer in dust) which can produce the same SED.  As an
example of this degeneracy, we fit the two galaxies NGC~4385 and
NGC~7714 for which we have complete SEDs from UV2 to K.  To test the
goodness of the fit for a specific starburst model, we used the
$\tilde{\chi}^2$ (reduced $\chi^2$) criterion after first normalizing
the observed and model SEDs at the K band.  The $\tilde{\chi}^2$ was
computed using
\begin{equation}
\tilde{\chi}^2 = \frac{1}{d} \sum_{i=1}^{n} \left(
    \frac{F_o(\lambda_i) - F_m(\lambda_i)}{\sigma(\lambda_i)}
    \right)^2
\end{equation}
where $n$ is the number of points in the observed SED, $d$ is the
number of degrees of freedom, $F_o(\lambda_i)$ is the observed flux at
the ith point in the SED, $F_m(\lambda_i)$ is the same for the model
flux, and $\sigma(\lambda_i)$ is the ith uncertainty in the observed
flux (\cite{tay82}).  The number of degrees of freedom $d = (n - 1)$
as both the observed and model SEDs were normalized at the K band.
Starburst models with $\tilde{\chi}^2 > 1.8$ were rejected as the
probability of such models agreeing with the observations was small.
With this limiting value of $\tilde{\chi}^2$, we will collect 96.3\%
of the starburst models which fit the observed SED within the
uncertainties (\cite{tay82}).  Such fits were obtained for 17 and 46
different starburst models for NGC~4385 and NGC~7714, respectively.

\begin{figure}[tbp]
\begin{center}
\plotone{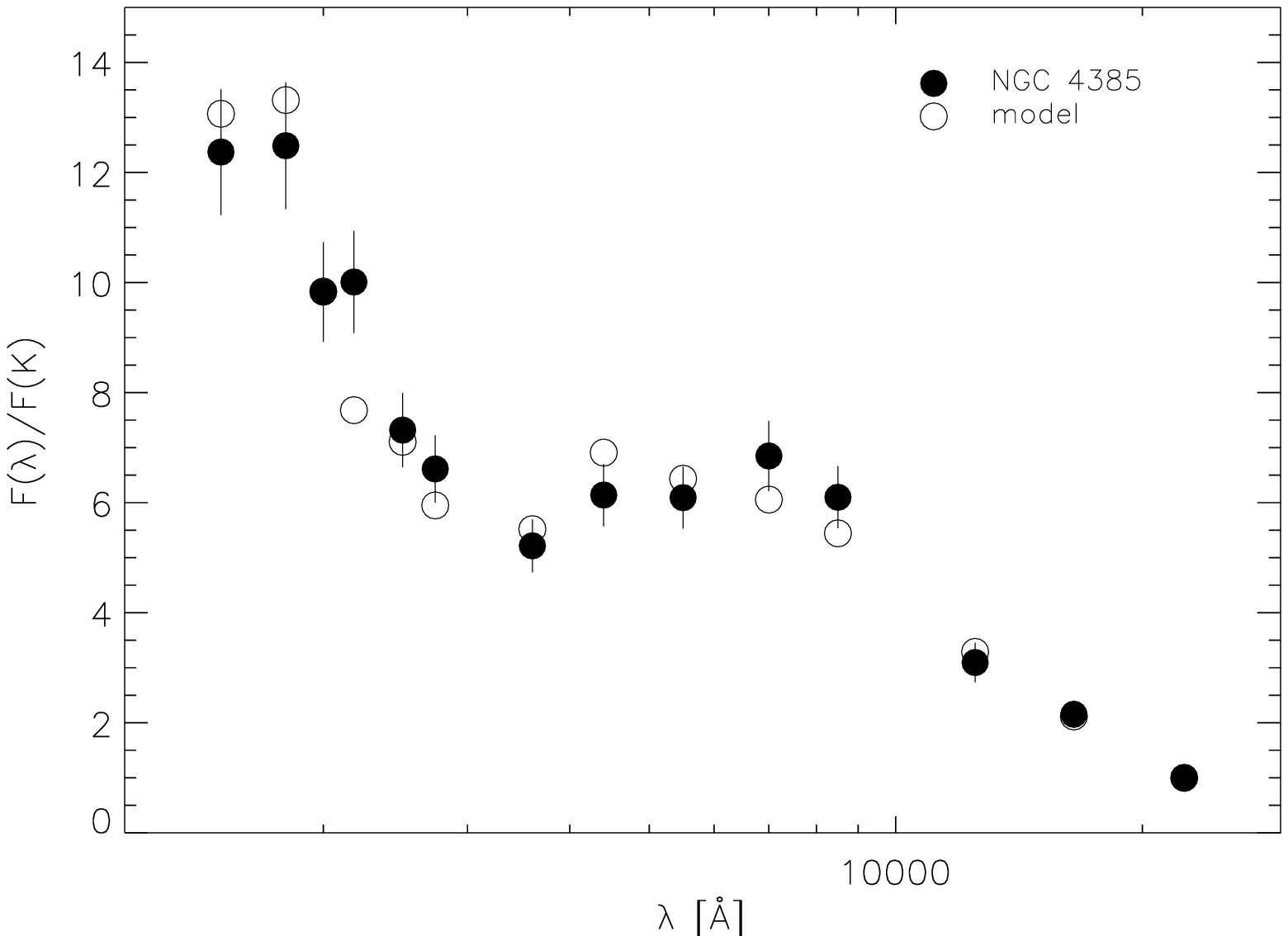} \\
\plotone{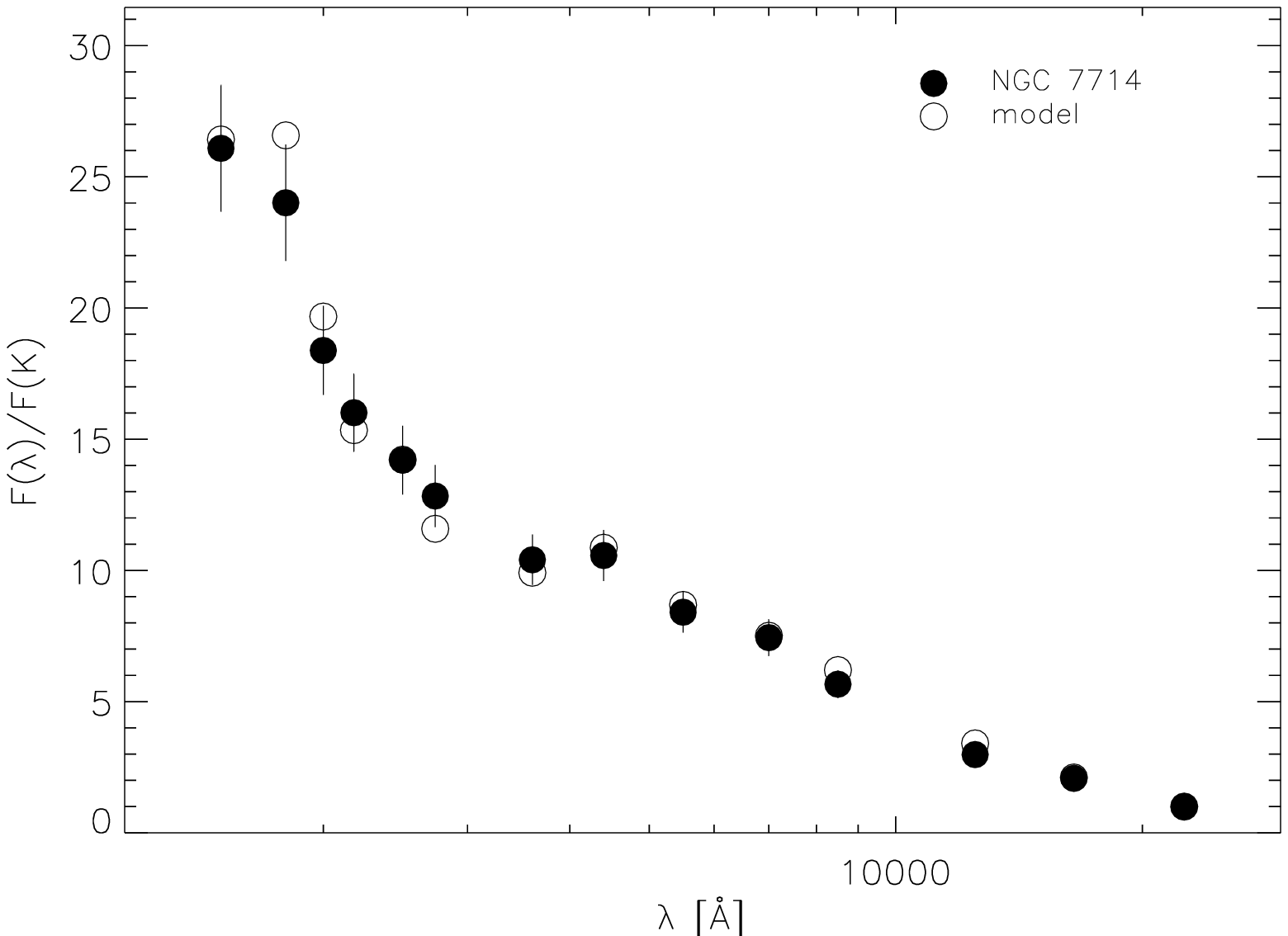}
\caption{The best fit starburst model for NGC~4385 (a, left) and
NGC~7714 (b, right) are plotted along with their observed
SEDs. \label{fig_fits}}
\end{center}
\end{figure}

  The parameters of the starburst models which fit are interesting. 
Only models containing dust with a SMC extinction curve fit the data,
consistent with our findings in $\S$\ref{sec_2175}.  Using our results
from $\S$\ref{sec_cc_plots} and restricting the fits to models
with a shell geometry, the number of starburst models which fit are 8
for NGC~4385 and 18 for NGC~7714.  In order to choose among the
starburst models which fit the observed SEDs, more information about
each starburst (e.g.\ the number of ionizing photons) must be used and
we plan to do this in a future paper.  The best fit starburst model
for NGC~4385 ($\tilde{\chi}^2 = 1.11$) was for a constant star
formation \cite{bru97} model with an age of $7 \times 10^9$~years, a
shell geometry with $\tau_V = 0.75$, and clumpy ($k_2/k_1 = 0.001$) SMC
type dust.  This best fit starburst model, along with the observed SED
of NGC~4385 is displayed in Figure~\ref{fig_fits}a.  The best fit
starburst model for NGC~7714 ($\tilde{\chi}^2 = 0.49$) was for a
constant star formation \cite{bru97} model with an age of $1 \times
10^9$~years, a shell geometry with $\tau_V = 1.00$, and clumpy
($k_2/k_1 = 0.001$) SMC type dust.  This best fit starburst model,
along with the observed SED of NGC~7714 is displayed in
Figure~\ref{fig_fits}b.

\begin{figure}[tbp]
\begin{center}
\plotone{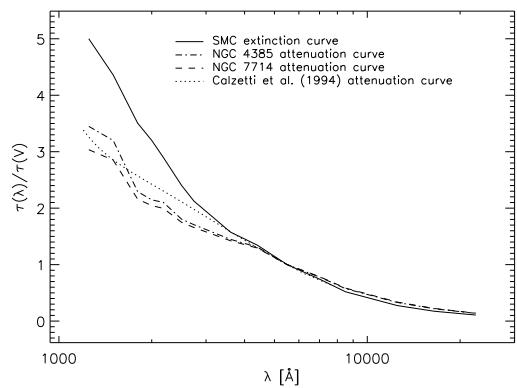}
\caption{This plot shows a comparison of the SMC extinction curve and
the Calzetti et al.\ (1994) average starburst attenuation curve with
the attenuations curves from the best fit starburst models of NGC~4385
and NGC~7714.  The average starburst attenuation curve was created by
normalizing the Calzetti et al.\ (1994) starburst attenuation curve to
the stellar continuum (E(B-V)~=~1) and adding ${\rm R}_{\rm V} = 3.1$.
The curves are normalized to the V band.  The effective $\tau_V$ for
NGC~4385 was 0.19 and the input $\tau_V$ was 0.75. The effective
$\tau_V$ for NGC~7714 was 0.25 and the input $\tau_V$ was 1.00.  The
effective $\tau_V$ is defined as the $\tau_V$ an overlaying screen
would need to produce the same reduction in flux.
\label{fig_atten_comp}}
\end{center}
\end{figure}

  To illustrate the difference between extinction and attenuation
curves, Figure~\ref{fig_atten_comp} displays the SMC extinction curve
and the average starburst attenuation curve (\cite{cal94}) with the
attenuation curves from the best fit starburst models for NGC~4385 and
NGC~7714.  For both galaxies, the attenuation curves are flatter than
the dust extinction curve and this is consistent with the Calzetti et
al.\ (1994) attenuation curve.  Also, the shape of the attenuation
curve is different between NGC~4385 \& NGC~7714 even though both
curves are for a shell geometry with a clumpy ($k_2/k_1 = 0.001$)
distribution of SMC dust.  This neatly shows the dependence of the
shape of an attenuation curve on both the geometry and amount of dust.
The systematic difference between the shape of both the NGC~4385 and
NGC~7714 attenuation curves and the Calzetti et al.\ (1994)
attenuation curve is likely due to differences between the actual dust
albedo and the adopted albedo (Table~\ref{table_dust_phys}).

  Calzetti et al.\ (1994) determined the effective B band optical
depth ($\tau_{\rm B,eff}$) to be 0.62 for NGC~4385 and 0.42 for
NGC~7714.  These values of $\tau_{\rm B,eff}$ were determined from
observations of H$\alpha$ and H$\beta$; therefore they measure the
$\tau_{\rm B,eff}$ of the starburst's gas.  The starburst models in
this paper measure the starburst's stars and from the best fit models
the $\tau_{\rm B,eff}$ was 0.26 for NGC~4385 and 0.33 for NGC~7714.
This result agrees with the finding by Calzetti (1997) that the gas in
starbursts shows a larger dust attenuation than the stars.  The gas is
likely associated with O stars (H~II regions), while the stellar
continuum has contributions from a wide range of stars.  As O stars
are short lived, on average they will be more deeply embedded in their
birth clouds and thus the gas will show a larger dust attenuation than
the stellar continuum.

\section{Discussion \label{sec_discussion}}

  The discovery that the dust in starbursts is closer to SMC type dust
than MW type dust is not totally unexpected (\cite{cal94};
\cite{mas96}).  It has been known that the attenuation curve of
starbursts is flat and does not show the presence of the 2175~\AA\
bump, but as a result of this work it is clear this is due to SMC-like
dust grains and not radiative transfer effects.  The 30~Dor region of
the LMC forshadows this conclusion as the dust associated with the
massive star formation in the 30~Dor region is very different than the
dust associated with the rest of the LMC.  The 30~Dor region dust has
a weaker 2175~\AA\ bump and stronger far-UV rise than the dust in the
rest of the LMC.  Like the 30~Dor region, it is unlikely the dust in
starbursts formed with a weak or negligible 2175~\AA\ bump and strong
far-UV rise.  It is more likely that the dust in starbursts has been
modified by the starburst activity to more SMC-like in nature.  Again,
the 30~Dor region has shown its identity as the Rosetta Stone for
starburst galaxies (\cite{wal91}).

  With this conclusion in hand the question arises: How does starburst
activity modify dust such that it loses its 2175~\AA\ bump and
strengthens its far-UV rise?  The answer lies in the different grain
populations which taken together explain dust extinction curves.  Dust
grains are thought to be composed of three different populations:
classically sized grains which produce the optical-IR extinction,
small grains which produce the 2175~\AA\ bump, and small grains which
produce the far-UV extinction rise (\cite{mat96}; \cite{zub96};
\cite{li97}).  The SMC-like extinction curve in starbursts implies the
2175~\AA\ bump grains are destroyed and the number of far-UV rise
grains have been increased.  The 30~Dor region has less intense star
formation leading to a similar but a notably different extinction
curve.  This extinction curve implies that the 2175~\AA\ bump grains
are still present, but in reduced numbers, and the far-UV rise grains
have increased their numbers.  Finally, the Orion region has a low
level of star formation ($\sim$1\% that of 30~Dor [\cite{hun95}])
which has also affected the dust's extinction curve.  The Orion region
extinction curve has a reduced 2175~\AA\ bump and a nearly {\em flat}
far-UV rise (\cite{car88}).  This implies both the 2175~\AA\ bump and
far-UV rise grains have been reduced in number.

  There seems to be a trend in the dust extinction curve with star
formation activity.  This can be explained by the relative importance
of shocks and stellar radiation in the modification of dust grains.
In the Orion region, the stellar radiation pressure and evaporation of
dust grains have selectively removed the small grains (\cite{mcc81};
\cite{car88}).  In starbursts, the dust modification is most likely
dominated by supernovae shocks which are efficient at shattering large
grains, thus increasing the small grain population (\cite{jon96}).
This explains the increase in the number of far-UV rise grains, but
the absence of the 2175~\AA\ bump grains implies they do not survive
the shocks and are either vaporized or shattered.  The 30~Dor region
lies between the Orion region and starburst galaxies in star formation
activity and, thus, has an extinction curve intermediate between the
two.  The trend with star formation activity implies the material
which makes up the 2175~\AA\ bump grains is not as robust as the
material which makes up the far-UV rise grains.  The identity of
material which gives rise to the far-UV rise and, especially, the
2175~\AA\ bump has remained elusive for many years.  For this reason,
we do not attempt to discuss the material which produces the 2175~\AA\
bump, but only to note that the trend with star formation activity of
the far-UV rise and 2175~\AA\ dust grains is a clue to the identity of
the material making those dust grains.

  The similarity of the average extinction curves of starbursts and
the SMC deserves comment.  The lack of a 2175~\AA\ bump and the steep
far-UV rise in the SMC extinction curve is attributed to the SMC's low
metallicity (\cite{pre84}; \cite{cla85}; \cite{fit86}).  Since our
sample of starbursts possess a large range of metallicities,
0.1--2~solar (see section~\ref{sec_obs}), this is unlikely to explain
the SMC-like extinction curve we found in starbursts.  We suggest the
SMC-like extinction curve in starbursts is a result of the massive
star formation.  Thus, both metallicity and star formation activity
may influence dust properties significantly.

  Observations which resolve starburst galaxies show that the star
formation occurs in localized regions analogous to the star clusters
and associations of our galaxy (\cite{oco94}; \cite{oco95};
\cite{whi95}; \cite{vac96}).  Not all starbursts seem to form stars
this way, see \cite{smi96} for an example of a global starburst.
These observations allow us to interpret the fact that we were able to
model the effects of dust in starbursts with a radiative transfer
model with a clumpy shell geometry.  The $\tau_V$ determined from our
starburst model is interpreted as the flux weighted average $\tau_V$
to many individual star clusters.  From this, it is not surprising we
find fairly low $\tau_V$'s as star clusters possessing a large
$\tau_V$ would not contribute much to the starburst SED.  The shell
geometry is a realistic approximation of the geometry of the
individual star clusters.  Near the center of the star clusters where
the radiation field is intense, dust grains would be unable to survive
and, thus, the inner region would be fairly free of dust.

  We were able to fit the SEDs of NGC~4385 \& NGC~7714 by using the
same radiative transfer model and only varying the amount of dust.
This particular model run (\#8, Table~\ref{table_mod_runs}) had SMC
dust, a shell geometry, a clumpy dust distribution with $k_2/k_1 =
0.001$.  This gives hope that it will be possible to investigate the
stellar evolutionary history of individual starbursts as well as the
type and amount of dust present.  These fits were done using only the
broad band SED of the starbursts and a proper fit would use the full
SED.  We plan to fit the full SED and fit both the continuum and
spectral lines in future work where we plan to investigate each
starburst individually.

\section{Conclusions \label{sec_conclusions}}

  From the comparison of 30 starburst SEDs with a model for starbursts
which included both stars and effects of dust, we concluded the
following:

\begin{enumerate}

\item The dust in starbursts has an extinction curve similar to that
found in the SMC, i.e.\ lacking a 2175~\AA\ bump and having a far-UV
rise with a slope intermediate between the MW and SMC.  The dust is
unlikely to have been formed with these properties, but to have
acquired them as a result of the massive star formation occurring in
the starburst.

\item The spatial distribution of stars and dust is best described by
a shell geometry.  This geometry has an inner dust-free sphere of
stars surrounded by a star-free shell of dust.

\item The dust distribution in starbursts is best described by a
locally clumpy distribution.

\item The age of the stellar populations in this UV selected sample of
starbursts ranges between $8 \times 10^{6}$ and $2 \times
10^{8}$~years if a burst star formation is assumed and $1 \times
10^{8}$ and $15 \times 10^{9}$~years if constant star formation is
assumed.

\item The effects of dust on the SED of a starburst are characterized
by an attenuation curve which is SMC-like extinction curve convolved
with the radiative transfer effects associated with a shell star/dust
geometry with a clumpy dust distribution.  The wanton use of a SMC
extinction curve to correct a starburst's SED is not supported by this
work.

\end{enumerate}

\acknowledgments

  We thank Sang-Hee Kim for computing the albedo and g values for the
SMC dust for us.  The comments of the referee, Geoff Clayton, were
appreciated and significantly improved this paper.  A major portion of
this work was done while K.\ D.\ Gordon was at the Space Telescope
Science Institute as a Collaborative Visitor of D.\ Calzetti.  The
Space Telescope Science Institute is operated by the Association of
Universities for Research in Astronomy, Inc., under NASA contract NAS
5-26555.  K.\ D.\ Gordon and A.\ N.\ Witt acknowledge financial
support from NASA LTSAP grants NAGW-3168 \& NAG5-3367 to The
University of Toledo.


\begin{thebibliography}{}
\bibitem[Bessell \& Brett (1988)]{bes88} Bessel, M.\ S.\ \& Brett, J.\
   M.\ 1988, \pasp, 100, 1134
\bibitem[Bessell 1990]{bes90} Bessel, M.\ S.\ 1990, \pasp, 102, 1181
\bibitem[Bianchi et al.\ 1996]{bia96} Bianchi, L., Clayton, G.\ C.,
   Bohlin, R.\ C., Hutchings, J.\ B., \& Massey, P.\ 1996, \apj, 471,
   203
\bibitem[Bohlin et al.\ 1990]{boh90} Bohlin, R.\ C., Harris, A.\ W.,
   Holm, A.\ V., Gry, C.\ 1990, \apjs, 73, 413
\bibitem[Bohlin 1996]{boh96} Bohlin, R.\ C.\ 1996, \aj, 111, 1743
\bibitem[Bohlin 1997]{boh97} Bohlin, R.\ C.\ 1997, private
   communication
\bibitem[Bruzual \& Charlot 1993]{bru93} Bruzual, A.\ G.\ \& Charlot,
   S.\ 1993, \apj, 405, 538
\bibitem[BC97]{bru97} Bruzual, A.\ G.\ \& Charlot, S.\ 1997, in
   preparation [BC97] 
\bibitem[Calzetti 1997]{cal97} Calzetti, D.\ 1997, \aj, 113, 162
\bibitem[Calzetti et al.\ 1995]{cal95} Calzetti, D., Bohlin, R.\ C.,
   Gordon, K.\ D., Witt, A.\ N., \& Bianchi, L.\ 1995, \apj, 466, L97
\bibitem[Calzetti et al.\ 1994]{cal94} Calzetti, D., Kinney, A.\ L.,
   \& Storchi-Bergmann, T.\ 1994, \apj, 429, 582
\bibitem[Calzetti et al.\ 1996]{cal96} Calzetti, D., Kinney, A.\ L.,
   \& Storchi-Bergmann, T.\ 1996, \apj, 458, 132
\bibitem[Cardelli \& Clayton 1988]{car88} Cardelli, J.\ A.\ \&
   Clayton, G.\ C.\ 1988, \aj, 95, 516
\bibitem[Cardelli, Clayton, \& Mathis 1989]{car89} Cardelli, J.\ A.,
   Clayton, G.\ C., \& Mathis, J.\ S.\ 1989, \apj, 345, 245
\bibitem[Clayton \& Martin 1985]{cla85} Clayton, G.\ C.\ \& Martin, P.\
   G.\ 1985, \apj, 288, 558
\bibitem[Dickey \& Garwood 1989]{dic89} Dickey, J.\ M.\ \& Garwood,
   R.\ W.\ 1989, \apj, 341, 201
\bibitem[Elias et al.\ 1982]{eli82} Elias, J.\ H., Frogel, J.\ A.,
   Matthews, K., \& Neugebauer, G.\ 1982, \aj, 87, 1029
\bibitem[Fitzpatrick 1985]{fit85} Fitzpatrick, E.\ L.\ 1985, \apj,
   299, 219
\bibitem[Fitzpatrick 1986]{fit86} Fitzpatrick, E.\ L.\ 1986, \aj, 92,
   1068
\bibitem[Fitzpatrick 1989]{fit89} Fitzpatrick, E.\ L.\ 1989, in IAU
   Symp.\ 135, Interstellar Dust, ed.\ L.\ J.\ Allamandola \& A.\ G.\
   G.\ M.\ Tielens (Dordrecht: Kluwer), 37
\bibitem[Fitzpatrick 1990]{fit90} Fitzpatrick, E.\ L.\ \& Massa, D.\
   1990, \apjs, 72, 163
\bibitem[Gordon et al.\ 1994]{gor94} Gordon, K.\ D., Witt, A.\ N.,
   Carruthers, G.\ R., Christensen, S.\ A., \& Dohne, B.\ C.\ 1994,
   \apj, 432, 641
\bibitem[Hoffleit \& Warren 1991]{hof91} Hoffleit, D.\ \& Warren, W.\
   H.\ Jr.\ 1991, The Bright Star Catalogue, 5th Revised Ed.\
   (Preliminary Version), Astronomical Data Center
\bibitem[Hunter et al.\ 1995]{hun95} Hunter, D.\ A., Shaya, E.\ J.,
   Holtzman, J.\ A., Light, R.\ M., O'Neil, E.\ J., \& Lynds, R.\
   1995, \apj, 448, 179
\bibitem[Jones, Tielens, \& Hollenbach 1996]{jon96} Jones, A.\ P.,
   Tielens, G.\ G.\ M., \& Hollenbach, D.\ J.\ 1996, \apj, 469, 740
\bibitem[Kim 1996]{kim96} Kim, S.-H.\ 1996, private communication
\bibitem[Kim, Martin, \& Hendry 1994]{kim94} Kim, S.-H., Martin, P.\
   G., \& Hendry, P.\ D.\ 1994, \apj, 422, 164
\bibitem[Kinney et al.\ 1993]{kin93} Kinney, A.\ L., Bohlin, R.\ C.,
   Calzetti, D., Panagia, N., \& Wyse, R.\ F.\ G.\ 1993, \apjs, 86, 5
\bibitem[Lan\c{c}on \& Rocca-Volmerange 1996]{lan96} Lan\c{c}on, A.\
   \& Rocca-Volmerange, B.\ 1996, New Astronomy, 1, 215
\bibitem[Lehtinen \& Mattila 1996]{leh96} Lehtinen, K.\ \& Mattila,
   K.\ 1996, \aap, 309, 570
\bibitem[Lequeux et al.\ 1982]{leq82} Lequeux, J., Maurice, E.,
   Pr\'evot-Burnichon, M.\ L., Pr\'evot, L., \& Rocca-Volmerange, B.\
   1982, \aap, 113, L15
\bibitem[LH95]{lei95} Leitherer, C.\ \& Heckman, T.\ M.\ 1995, \apjs,
   96, 9 [LH95] 
\bibitem[Li \& Greenberg 1997]{li97} Li, A.\ \& Greenberg, J.\ M.\
   1997, \aap, in press
\bibitem[Maeder \& Conti 1994]{mae94} Maeder, A.\ \& Conti, P.\ S.\
   1994, \araa, 32, 227
\bibitem[Mas-Hesse \& Kunth 1996]{mas96} Mas-Hesse, J.\ M.\ \& Kunth,
   D.\ 1996, in The Interplay Between Massive Star Formation, The ISM
   and Galaxy Evolution, eds.\ D.\ Kunth et al.\ (France: Editions
   Fronteires), 401
\bibitem[Massa, Savage, \& Fitzpatrick 1983]{mas83} Massa, D., Savage,
   B.\ D., Fitzpatrick, E.\ L.\ 1983, \apj, 266, 662
\bibitem[Mathis 1996]{mat96} Mathis, J.\ S.\ 1996, \apj, 472, 643
\bibitem[Mathis \& Cardelli 1992]{mat92} Mathis, J.\ S.\ \& Cardelli,
   J.\ A.\ 1992, \apj, 398, 610
\bibitem[McCall 1981]{mcc81} McCall, M.\ L.\ 1981, \mnras, 194, 485
\bibitem[Mcquade, Calzetti, \& Kinney 1995]{mcq95} Mcquade, K.,
   Calzetti, D., \& Kinney, A.\ L.\ 1995, \apjs, 97, 331
\bibitem[O'Connell, Gallagher, \& Hunter 1994]{oco94} O'Connell, R.\
   W., Gallagher, J.\ S.\ III, \& Hunter, D.\ A.\ 1994, \apj, 433, 65
\bibitem[O'Connell et al.\ 1995]{oco95} O'Connell, R.\ W., Gallagher,
   J.\ S.\ III, Hunter, D.\ A., \& Colley, W.\ N.\ 1995, \apj, 446, L1 
\bibitem[Pr\'evot et al.\ 1984]{pre84} Pr\'evot, M.\ L., Lequeux, J.,
   Maurice, E., Pr\'evot, L., \& Rocca-Volmerange, B.\ 1984, \aap,
   132, 389
\bibitem[Rosen \& Bregman 1995]{ros95}  Rosen A.\ \& Bregman,
   J.\ N.\ 1995, \apj, 440, 634  
\bibitem[Salpeter (1955)]{sal55} Salpeter, E.\ E.\ 1955, \apj, 121,
   161
\bibitem[Scalo 1990]{sca90} Scalo, J.\ 1990, in Physical Processes in
   Fragmentation \& Star Formation, eds.\ R.\ Capuzzo-Dolcetta et
   al.\ (Dordrecht: Kluwer)
\bibitem[Smith et al.\ (1996)]{smi96} Smith, D.\ A., et al.\ 1996, \apj,
   473, L21
\bibitem[Storchi-Bergmann, Calzetti, \& Kinney 1994]{sto94}
   Storchi-Bergmann, T., Calzetti, D., \& Kinney, A.\ L.\ 1994, \apj,
   429, 572
\bibitem[Storchi-Bergmann, Kinney, \& Challis 1995]{sto95}
   Storchi-Bergmann, T., Kinney, A.\ L., \& Challis, P.\ 1995, \apjs,
   98, 103
\bibitem[Taylor 1982]{tay82} Taylor, J.\ R.\ 1982, An Introduction to
  Error Analysis (Mill Valley, CA: Univ.\ Science Books)
\bibitem[T\"ug, White, \& Lockwood 1977]{tug77} T\"ug, H.,
   White, N.M., \& Lockwood, G.W. 1977, \aap, 61, 679
\bibitem[Vacca 1996]{vac96} Vacca, W.\ D.\ 1996, in The Interplay
   Between Massive Star Formation, The ISM and Galaxy Evolution, eds.\
   D.\ Kunth et al.\ (France: Editions Fronteires), 321
\bibitem[Walborn 1991]{wal91} Walborn, N.\ R.\ 1991, in Massive Stars
   in Starbursts, eds.\ C.\ Leitherer et al.\ (Cambridge: Cambridge
   Univ. Press), 145
\bibitem[Wesselius et al.\ 1982]{wes82} Wesselius, P.\ R., van Duinen,
   R.\ J., de Jonge, A.\ R., Aalders, J.\ W.\ G., Luinge, W., \&
   Wildeman, K.\ J.\ 1982, \aaps, 49, 427
\bibitem[Whitmore \& Schweizer 1995]{whi95} Whitmore, B.\ C.\ \&
   Schweizer, F.\ 1995, \aj, 109, 960
\bibitem[Whittet 1992]{whi92} Whittet, D.\ C.\ B.\ 1992, Dust in the
   Galactic Environment (Bristol: IOP)
\bibitem[Witt 1977]{wit77} Witt, A.\ N.\ 1977, \apjs, 35, 1
\bibitem[Witt, Bohlin, \& Stecher 1984]{wit84} Witt, A.\ N., Bohlin,
   R.\ C., \& Stecher, T.\ P.\ 1984, \apj, 279, 698
\bibitem[Witt \& Gordon 1996]{wit96} Witt, A.\ N.\ \& Gordon, K.\ D.\
   1996, \apj, 463, 681
\bibitem[Witt, Gordon, \& Madsen (1997)]{wit97} Witt, A.\ N., Gordon,
   K.\ D., \& Madsen, G.\ M.\ 1997, in preparation
\bibitem[WTC]{wit92} Witt, A.\ N.,
   Thronson, H.\ A., \& Capuano, J.\ M.\ 1992, \apj, 393, 611 [WTC]
\bibitem[Worthey 1994]{wor94} Worthey, G.\ 1994, \apjs, 95, 107
\bibitem[Zubko, Kre\l{}owski, Wegner 1996]{zub96} Zubko, V.\ G.,
   Kre\l{}owski, J., \& Wegner, W.\ 1996, \mnras, 283, 577
\end{thebibliography}
\end{document}